# Multipacting Analysis in Micro-pulse Electron Gun Design*

Liao Lang(廖浪)[1,2;2]    Zhang Meng(张猛)[1]    Gu Qiang(顾强)[1]    Fang Wen-Cheng(方文程)[1]

Zhao Ming-Hua(赵明华)[1;1]

[1] Shanghai Institute of Applied Physics, Chinese Academy of Sciences, Shanghai 201800, China

[2] Graduate University of Chinese Academy of Sciences, Beijing 100049, China

**Abstract:** Modeling multipacting to steady state saturation is of interest in determining the performance of micro-pulse electron gun. In this paper, a novel method is proposed to calculate the multipacting resonance parameters for the gun. This method works well, and the 2-D simulation results suggest that steady state saturation can be achieved in the gun. After saturation the transition from two-surface multipacting to single-surface multipacting is occurred, and an extensive range of electrons emission time is a suggested way to avoid this kind of transition.

## 1    Introduction

In recent decades, the phenomenon of multipacting has been deeply investigated in many areas in order to suppress its adverse effects[1]. The Micro-Pulse Electron Gun (MPG), which as a positive application of multipacting effect, can be triggered with a single electron and resonantly amplified to saturation inside a vacuum RF cavity[2]. MPG can give rise to electrons beam with high currents and short pulses, which is useful for industrial applications as well as high energy accelerators[3].

Owing to the complexity of multipacting effect and its great impact on the performance of MPG, study should be focused in detail. The main purpose of this paper is the description of one novel method for determining the resonance condition in MPG design. The design model consists of a reentrant cavity and a grid-anode (as shown in Fig. 1), across which a time varying RF voltage is applied. Under the condition of resonance, steady state saturation is achieved in 2-D simulations. In addition, the transition from two-surface multipacting to single-surface multipacting, which found to prevent MPG from outputting micro-pulses, is occurred after saturation in 3-D simulations, and we obtained a solution to suppress this kind of transition.

The article is organized as follows: In Sec.2, we first present the MPG concept, and a novel solution which is used to calculate resonance parameters easily for MPG design is provided based on the basic equations. In Sec.3, 2-D VORPAL simulations have been taken in order to confirm this kind of method. In Sec.4, 3-D simulation results are provided for the design, and finally we give a brief conclusion.

---

*Supported by Major State Basic Research Development Program of China (973 Program) (Grant No. 2011CB808300) and National Natural Science Foundation of China (Grant No. 10935011)

1) Corresponding author, E-mail: zhaominghua@sinap.ac.cn

2) E-mail: liaolang@sinap.ac.cn

## 2 A novel method to calculate resonance parameters

### 2.1 MPG concept and electron resonance equation

The micro-pulse electron gun is one kind of microwave gun that employs secondary electrons which are resonantly amplified in a RF cavity to produce micro-pulses[1]. A number of initial electrons, which are emitted off one side (as shown in Fig. 1), are transited the cavity under the work of RF field, and finally they struck the other side in odd multiple of half period and generated secondary electrons.

In the gun, the multiple of secondary electrons in one period can be written as $\delta_1\delta_2(1-T)$[2]. In order to have a gain in the cavity, following condition should be met:

$$\delta_1\delta_2(1-T) > 1. \tag{1}$$

where $\delta_1, \delta_2$ are the secondary emission yield of cathode and grid-anode. T is the transmission factor of the grid-anode.

The effect of multipacting in MPG is similar to that in the parallel-plate, and the equation of motion of single electron is:

$$d^2y/dt^2 = (eV_{rf0}/mD)\sin(\varphi+\omega t). \tag{2}$$

where $V_{rf0}$ is the amplitude of RF voltage and $\omega=2\pi f$ (f is the frequency of cavity).

Assuming that electrons had an initial velocity $v_0$ at position y=0 with an initial phase $\varphi=\theta_0$, then velocity dy/dt and position y(t) of the particle are readily solved from Eq. (2) to be

$$dy/dt = \frac{eV_{rf0}}{mD\omega}\left[\cos\theta_0 - \cos(\varphi+\omega t)\right] + v_0, \tag{3}$$

$$y = \frac{eV_{rf0}}{mD\omega^2}\left[\sin\theta_0 - \sin(\varphi+\omega t) + \omega t\cos\theta_0\right] + v_0. \tag{4}$$

When the electrons reach the opposite side at y=D, the transit-time $\omega t = N\pi$ (N is an odd number), and $\varphi=\theta_0$. So the impact velocity $v_y$ and resonance equation can be written as follows:

$$v_y = \frac{2eV_{rf0}}{m\omega D}\cos\theta_0 + v_0, \tag{5}$$

$$V_{rf0} = \frac{m\omega D(\omega D - N\pi v_0)}{e(N\pi\cos\theta_0 + 2\sin\theta_0)}. \quad (N=1,3,5\ldots) \tag{6}$$

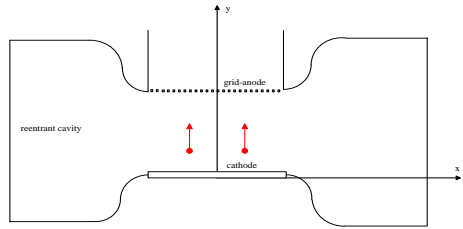

Fig. 1.   The reentrant cavity of micro-pulse electron gun.

### 2.2 A novel method for MPG design

It is convenient to find a resonance condition by using Eq. (5) and (6). The steps to figure out the basic parameters have been shown in the Fig. 2. Here, we set the initial $\theta_0$ to 0 for the sake of simplicity.

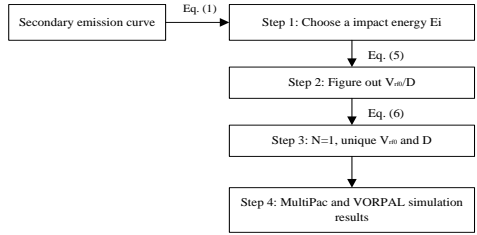

Fig. 2. A flow chart used to calculate the resonance parameters in the cavity.

In step 3, we set N to 1 for two reasons: Firstly the first-order multipacting has a wider resonance voltage range than that for higher orders[4]; The second reason is that the first-order multipacting can give rise to better pulses quality. Depending on electrons emission phase, hybrid-orders mode is commonly existent in multipacting as shown in Fig. 3. Setting N to 1 is a suggested solution to avoid the possibility of hybrid-orders. Based on this setting, the resonance parameters are calculated as shown in the Table 1.

Table 1. Basic simulation condition

| Cathode/Anode Materials | Frequency / GHz | Target Impact Energy / KeV |
|---|---|---|
| MgO | 2.856 | 5 |
| Gap Distance D / mm | $V_{rf0}/D$ / (MV/m) | Transmission of grid-anode |
| 4 | 2.08 | 50% |

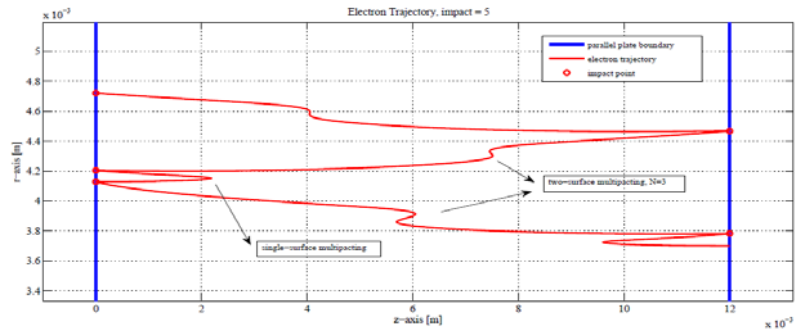

Fig. 3. Hybrid-orders mode in multipacting, for gap length D=1.2cm.

MultiPac, which is widely used by researchers around the world, is a 2D multipacting code with known secondary emission properties[5]. As gap distance D and RF electric field $V_{rf0}$ have been calculated, scanning an extensive electric field zone in MultiPac is proposed here in order to find a proper resonant condition, and Fig. 4 shows the simulation results.

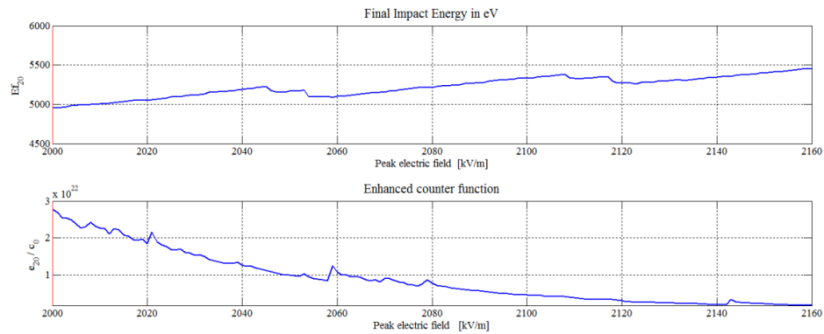

Fig. 4. MultiPac simulation results. (above)Average impact energy in eV after 20 impacts; (below)Enhance function after 20 impacts, note that multipacting occurs only when the enhanced function is greater than one. MultiPac neglects the space charge effect.

In short conclusion, this method provide a convenient way to determine the resonance condition in the gun. Combing with the MultiPac results, the proper running parameters for the gun is gotten, and these parameters work as the input in our further simulations.

## 3  2-D VORPAL simulation

VORPAL is the PIC simulation code for complex geometries and allowing the study of effects of multipacting in vacuum devices[6-7]. It is easier for VORPAL to simulate the effect of multipacting with up to hundreds billion of physical electrons than other codes, such as CST. Table 1 shows the basic parameters for both 2-D and 3-D simulations. At the beginning, a y-direction sinusoidal RF voltage has been motivated in the cavity, a group of electrons have been loaded ten periods later in order to synchronize with the field. The electrons emitted from the cathode and hit the grid-anode a half of period later, and the process of grid-hitting can't be directly observed before saturation in Fig. 8, but it exactly showed as a sub-peak of curve after saturation.

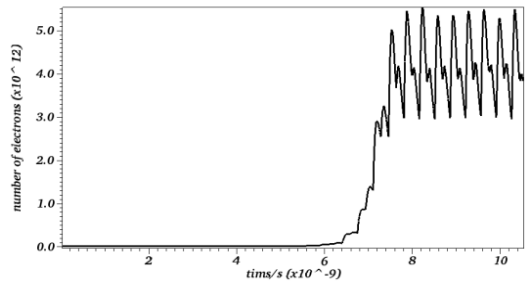

Fig. 8.   Number of physical electrons as a function of time in 2-D simulation, steady state saturation is occurred after 8ns.

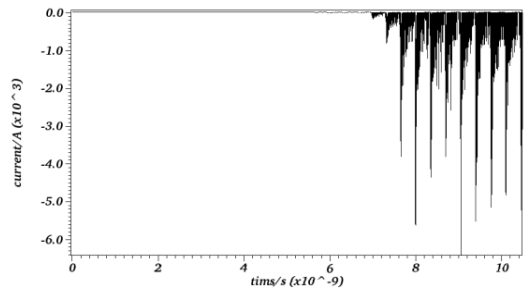

Fig. 9.   The time history of currents monitored by a sheet behind the grid-anode.

Steady state saturation, which is need for the running of MPG, is occurred after 8ns. The Fig. 9 shows the currents as a function of time, and the currents that monitored outside of the cavity saturates at about 5500 A.

## 4  A transition phenomenon from two-surface multipacting to single-surface multipacting in MPG

The final steady state characteristics of multipacting discharge are important for MPG. Steady state saturation is achieved in our 2-D VORPAL simulation, and it is necessary to construct a 3-D model to study multipacting effects because of its complexity. Here TM010 has been motivated in the 3-D reentrant cavity, and steady state of $E_z$ occurred ten periods later (as shown in Fig. 10). Table 2 shows two kinds of codes for running VORPAL simulations, the conditions are same except for a different emission time.

As the same with the reference [8], the transition that caused by space charge effect from two-surface multipacting to single-surface multipacting in MPG is observed by analysis of simulation results. In Fig. 11, the impact currents of grid-anode gave rise to peak due to saturation, then it gradually decreased with the axial electric field and finally to zero, while the cathode currents still kept steady state. Those changes strongly suggest the existence of transition from two-surface multipacting to single-surface multiacting.

Table 2.    electrons emission time for VORPAL code

| code | start_emission | end_emission |
|---|---|---|
| 1 | 10*period | 10*period+1*period |
| 2 | 10*period | 10*period+5*period |

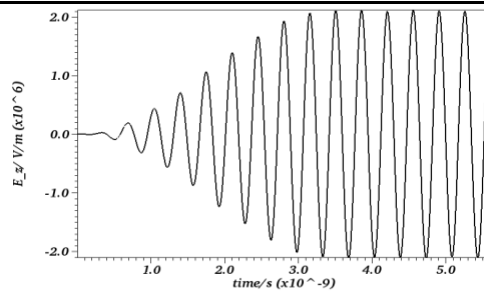

Fig. 10.   Z-direction electric field E_z as a function of time.

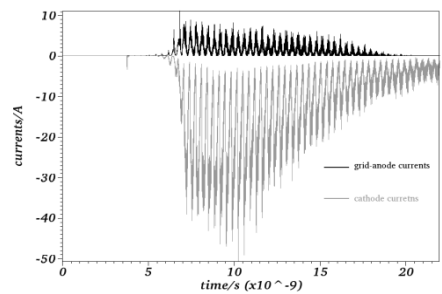

Fig. 11. Time history of cathode and grid-anode currents (namely, monitored in the cathode and grid-anode) for code 1 in Table 2.

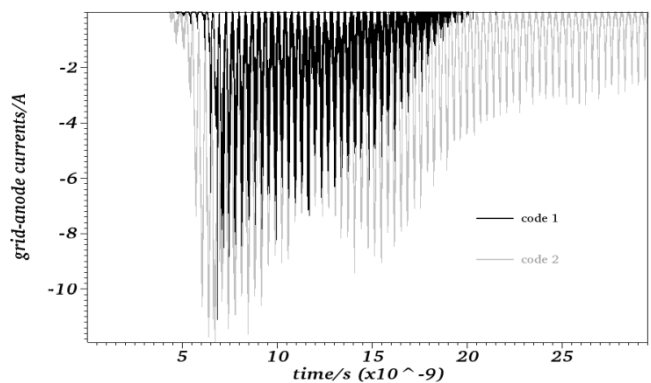

Fig. 12.   Currents monitored in grid-anode. The anode impact currents of code 1 (light dark curve) is compared with impact currents of code 2 (darker shading curve). The darker shading curve decreased to 0 after 18ns, suggesting no impacts in the grid-anode. The light dark curve indicates steady state saturation, the time distance of nearby two summits is one period.

The transition from two-surface multipacting to single-surface multipacting would destruct the mechanism of micro-pulse gun and it is not expected in our gun design. This kind of transition caused by the mismatch between electrons emission time and resonance voltages. Enhancing the range of emission time would help to determine the resonance condition. The light dark curve in Fig. 12 shows the solution to suppress the transition, and the currents after 25ns indicate a steady state for multiapcting.

## 5   Conclusion

In this paper, a comprehensive analytical study has been taken about the effect of multipacting in micro-pulse electron gun with the combination of 2-D and 3-D simulations. The basic parameters for cavity of gun had been calculated, it work as the input parameters for MultiPac and VORPAL. In 2-D VORPAL simulation, steady state saturation has been reached due to the balance of electrons loss caused by space charge effect and secondary emission, but a harmful transition from two-surface multipacting to single-surface multipacting found in a few VORPAL 3-D simulations would prevent micro-pulse gun from outputting micro-pulses. In case of this kind of transition, an extensive electrons emission time has been used in order to have a steady performance for the gun.

*We thank Dazhang Huang for instructive discussions and help. We would also like to thank Hongyu Wang, Chuandong Zhou for their assistances in VORPAL's use.*

## References

1   J M. Vaughan. IEEE TRANSACTIONS ON ELECTRON DEVICES, 1988, **36**: 1172

2   F. Mako. IEEE, 1993. 2702-2704

3   L. K. Len, F. Mako. Self-Bunching Electron Gun. Proc of PAC99. IEEE, 1999. 70-74

4   Rami Alfred Kishek. Interaction of Multipactor Discharge and RF Structures (Ph. D. Thesis). University of Michigan, 1997

5   D. Naik, Ilan Ben-Zvi. Ripple Structure In A 56MHz Quarter Wave Resonator for Multipacting Suppression. Proc of PAC09. 2009. 903-905

6   C. Nieter, John R. Cary. Journal of Computational Physics, 2004. 448-473

7   C. Nieter, John R. Cary, Gregory R. Werner et al. Journal of Computational Physics, 2009. 7902-7916

8   C. J. Lingwood, G. Burt, A. C. Dexter et al. Physics of Plasmas, 2012, **19**: 032106